\documentclass[onecolumn]{aastex631}

\usepackage{graphicx}	
\usepackage{amsmath}	

\shorttitle{GRB jet opening angle}
\shortauthors{Xu et al.}
\begin{document}

\title{Constrain the GRB jet opening angle based on the very steep decay phase}


\correspondingauthor{Yuan-Chuan Zou}
\email{zouyc@hust.edu.cn}

\author{Qian Xu}
\affiliation{School of Physics, Huazhong University of Science and Technology, Wuhan 430074, China\\}
\author{Dong-Jie Liu}
\affiliation{School of Physics, Huazhong University of Science and Technology, Wuhan 430074, China\\}
\author[0000-0002-5400-3261]{Yuan-Chuan Zou}
\affiliation{School of Physics, Huazhong University of Science and Technology, Wuhan 430074, China\\}
\affiliation{Purple Mountain Observatory, Chinese Academy of Sciences, Nanjing, 210023, China}

\begin{abstract}
Thanks to the rapid follow-up observations by \textit{Swift}/XRT, a good part of Gamma-Ray Bursts (GRBs) high latitude emission have been observed in X-ray band.
Some of them even show a dropdown decay after this period, which strongly indicates the edge of the jet is corresponding to the breaking time.
This study constrains the jet opening angles of GRBs by analyzing the very steep decay phase in the early X-ray afterglow. 
Using data from \textit{Swift}/XRT, we identified GRBs with significant breaks in their light curves and applied a broken power-law model to describe the decay phases.
Assuming a spherical and isotropic emitting surface, we set constraints on the radiation radius ($R_{\gamma}$) to estimate jet opening angles ($\theta_{\rm jet}$) from the breaking time.
Our results indicate that jet opening angles can be constrained, although they are sensitive to the assumed radiation radius. This approach provides yet another method for estimating GRB jet opening angles.
\end{abstract}

\keywords{gamma-ray burst: general -- stars: jets -- relativistic processes}



\section{Introduction}
Gamma-ray bursts (GRBs) are among the most energetic explosions in the universe, releasing tremendous amounts of energy across the entire electromagnetic spectrum over a brief period of time. These intense bursts of gamma rays are believed to originate from the collapsed remnants of massive stars (long GRBs) or the merging of compact object binaries such as neutron stars or black holes (short GRBs; see \citealt{2004RvMP...76.1143P, 2015PhR...561....1K} for reviews).

A key characteristic of GRBs is the existence of narrowly collimated relativistic jets produced during the explosion \citep{Rhoads1999, Sari1999}. The ultra-relativistic nature of these jets, with Lorentz factors typically in the range of 100-1000, is required to overcome the ``compactness problem" and allow the emission of gamma rays \citep{Lithwick2001}. The jet's high bulk Lorentz factor ($\Gamma$) also helps explain the non-thermal spectra observed in GRBs via particle acceleration in internal shocks within the jet \citep{Rees1994, Daigne1998}.
In addition to the Lorentz factor, the opening angle ($\theta_{\rm jet}$) of the GRB jet plays a crucial role in understanding the true energetics and event rates of these phenomena \citep{Frail2001, Bloom2003}.

One of the primary methods to constrain the jet opening angle is through the observation of a ``jet break" in the afterglow light curve. This jet break occurs when the relativistic beaming angle ($1/\Gamma$) becomes comparable to the jet opening angle ($\theta_{\rm jet}$), causing a steepening in the light curve \citep{Rhoads1999, Sari1999}. 
Observations of such jet breaks in the afterglows of GRBs have provided estimates of $\theta_{\rm jet}$, leading to important insights into the true energy scale of GRBs \citep{Frail2001, Bloom2003}.

The high latitude emission may emerge once the emitting shell ceases in the prompt phase. 
It has been observed as a steep decay phase in the early stage of the X-ray emission \citep{Zhang2006}. 
This happens when the photons emit from the high latitude with angle $1/\Gamma < \theta < \theta_{\rm jet}$. 
With later time, the emitting region comes from higher latitude $\theta$.
Once it reaches the edge of the jet ($\theta_{\rm jet}$), there should be an abrupt drop in the light curve as no more emission from even larger $\theta$.
We call it the dropdown decay phase. 
There are quite a few GRBs show such property with the decaying index down to -10. 

In this work, we concentrate on this scenario. 
Because of the rapid response of XRT on board {\it Swift} satellite \citep{Gehrels2004}, the high latitude emission can be seen on many GRBs with XRT observations \citep{Zhang2006}. Some of them show the dropdown decay after the steep decay. This break indicates the edge of the jet, and the opening angle can be inferred in principle.
We introduce the details of the model in Section \ref{sec:Model}, the data and results are given in Section \ref{sec:Sample}, and the conclusions and discussions are given in section \ref{sec:Con}.

\section{Model}\label{sec:Model}
Based on the internal shock model of the GRB emission, the prompt $\gamma$-rays are emitted from the shocked shells. However, the higher latitude emission may also reach to the observer at later times, and the frequency of the emission is lower as Doppler transformation effect. This shows as a steep decaying phase in the early X-ray afterglow \citep{Zhang2006}. 
The decay rate $\alpha$ obeys the so-called curvature effect relation $\alpha=2+\beta$, where $\beta$ is the spectral index, i.e., $F_\nu \propto \nu^{-\beta} t^{-\alpha}$ \citep{2000ApJ...541L..51K}.
Considering the opening angle of the jet, once after the time $t_{\rm jet}$ the emission comes from the edge of the jet, there is no more emission can be seen from an even higher latitude. 
The emission drops abruptly. 
Other emission such as the cooling tail of the emitting region may take over the decay, and we see a dropdown ($>2+\beta$) decay in the light curve. 
Once we find the breaking time $t_{\rm jet}$, the opening angle of the jet can generally be determined.

The scenario is based on the following assumption.
First, we assume that the emitting surface at radius $R_\gamma$ is spherical and that the emission is isotropic in the comoving frame. 
Secondly, the jet is assumed to pointing to the observer, which is $\theta_{\rm obs} \sim 0$. This is roughly true, otherwise, the prompt emission of the GRBs may not be seen.
Thirdly, other origins of the emission, such as the Electron-positron pair-enriched ejecta, reverse shock emission, and the energy injection \citep{2009MNRAS.395..955B} are not considered. Otherwise, there may not exist a dropdown phase. That means, this method is not suitable for those GRBs for constraining the opening angle. We will just omit those GRBs in this work.

The breaking time from steep decay to dropdown decay is related to the jet opening angle as follows \citep[see][for example]{2012A&A...542L..29H},
\begin{equation}
   {t_{\rm jet}} = \frac{R_{\gamma} \theta_{\rm jet}^2}{2 c}.
    \label{equ_1}
\end{equation}
The breaking time can solely determine the combination $R_{\gamma} \theta_{\rm jet}^2$. If we want to get the opening angle, $R_{\gamma}$ should be determined first. However, the most commonly used relation $R_\gamma = 2 \Gamma^2 c \delta T$, where $\Gamma$ is the Lorentz factor of the internal shocked jet, and $\delta T$ is the timescale of the prompt light curve, is model dependent and parameter dependent. Here, we use the upper limit and the lower limit of $R_\gamma$ to make constraints on the opening angle $\theta_{\rm jet}$.

We set the lower limit based on the compactness problem, which is that the high-energy photons should be absorbed by the lower energetic photons if the radius $R_\gamma$ is too small. Though, it is often used to constrain the lower limit of the Lorentz factor of GRB jet \citep{1993A&AS...97...59F}. 

The other constraint on the lower limit is that the $\gamma-$ray photons cannot escape if the optical depth is greater than 1. This induces the radius should be greater than \citep[see equation (27) in][]{1999PhR...314..575P}.
\begin{equation}
    R_{\rm e} \simeq 6 \times 10^{13} E_{52}^{1/2} \Gamma_{2}^{-1} {\rm cm},
    \label{equ_2}
\end{equation}
where $E$ is the isotropic equivalent kinetic energy of the GRB. 
$Q_k = Q/10^k$ is used in the whole paper. All quantities are in cgs units. 
Based on both constraints, we choose $10^{14}$ cm as the lower limit of the radiation radius.

For the upper limit, we set it being the deceleration radius $R_{\rm dec}$. If $R_\gamma$ is larger than $R_{\rm dec}$, most energy should have been released through the deceleration by the circum-burst medium, and the radiation is afterglow-like, i.e., the light curve is smoother compared with the very variable prompt emission. The deceleration radius for uniform circum-burst density is \citep[see equatoin (7.79) in][]{2019pgrb.book.....Z}
\begin{equation}
    R_{\rm dec} \simeq 6.2 \times 10^{16} E_{52}^{1/3} \Gamma_{2}^{-2/3} n^{-1/3} {\rm cm},
    \label{equ_3}
\end{equation}
where $E$ is the isotropic equivalent kinetic energy of the GRB, and $n$ is the number density of the circum-burst medium. Considering the typical $E$ is in order of $10^{54}$ ergs for a whole GRB, $E \sim 10^{52}$ ergs is a proper typical isotropic equivalent kinetic energy of a certain shell. The initial Lorentz factor $\Gamma \sim 100$ is a typical value. And the number density $n \sim 1 {\rm cm}^{-3}$ is also a typical value. Here, we set $R_{\rm dec} = 10^{17}$ cm as a typical upper limit of $R_\gamma$. Notice for individual GRB, the upper limit could vary (not much) depending on the energy, the circum-burst density etc. 

Once the $t_{\rm jet}$ is obtained, one can get the lower and upper limits of $\theta_{\rm jet}$ based on the upper and lower limits of $R_{\gamma}$ given above.

\section{Sample and Results}\label{sec:Sample}

The data are selected from the light curves of Swift/XRT \footnote{\url{https://www.swift.ac.uk/xrt_curves/allcurves.php}}, with GRBs starting from GRB 041223 until GRB 231230A. In total, there are, 1422 GRBs.
We visually inspected the light curves of each GRB and selected those bursts with noticeable break in the early afterglow as our preliminary data samples, as the light curve with no break can be clearly ruled out. Then we fit the light curves and classified them into golden sample, silver sample and those not included based on the criteria described later on in this section. 

The behavior patterns before and after the jet break\footnote{Notice this jet break is different from the commonly called jet break during the late afterglow stage, though they are both breaks caused by jet.} can be described using a broken power-law function:
\begin{equation}
    \left.f(t)=\left\{\begin{array}{lr}A(t/t_{\rm jet})^{-\alpha_1}, &t<t_{\rm jet}, \\
    A(t/t_{\rm jet})^{-\alpha_2}, &t>t_{\rm jet}, \end{array}\right.\right.
\end{equation}
where $A$ is a normalization term, $\alpha_1$ and $\alpha_2$ represent the decay indices before and after the jet break, and $t_{\rm jet}$ is the breaking time. For each GRB, the starting time ($t_{\rm start}$) and ending time ($t_{\rm end}$) of the two decaying stages before and after the jet break are determined through visual inspection. 
Then, data within this period are selected for fitting. 
The jet break time $t_{\rm jet}$ can be determined through the fitting.
Given the time range [$t_{\rm start}$, $t_{\rm jet}$], the spectral index ($\beta_1$) before the jet break can be obtained by using the data supplied by the UK Swift Science Data Centre at the University of Leicester \citep{Evans2009}.
Once the closure relation $\alpha_1=\beta_1+2$ is satisfied, we confirm the decay is the high latitude emission and the fitted $t_{\rm jet}$ is the jet break time.
Then $R_{\gamma} \theta_{\rm jet}^2$ can be determined through Equation (\ref{equ_1}).
As discussed above in Equations (\ref{equ_2}) and  (\ref{equ_3}), we choose $10^{14}$ cm and $10^{17}$ cm as the lower and upper bounds of $R_{\gamma}$. 
Consequently, the lower and upper limits for the jet opening angle $\theta_{\rm jet}$ are obtained through Equation (\ref{equ_1}).

\begin{figure}
    \centering
    \includegraphics[width=0.5\textwidth]{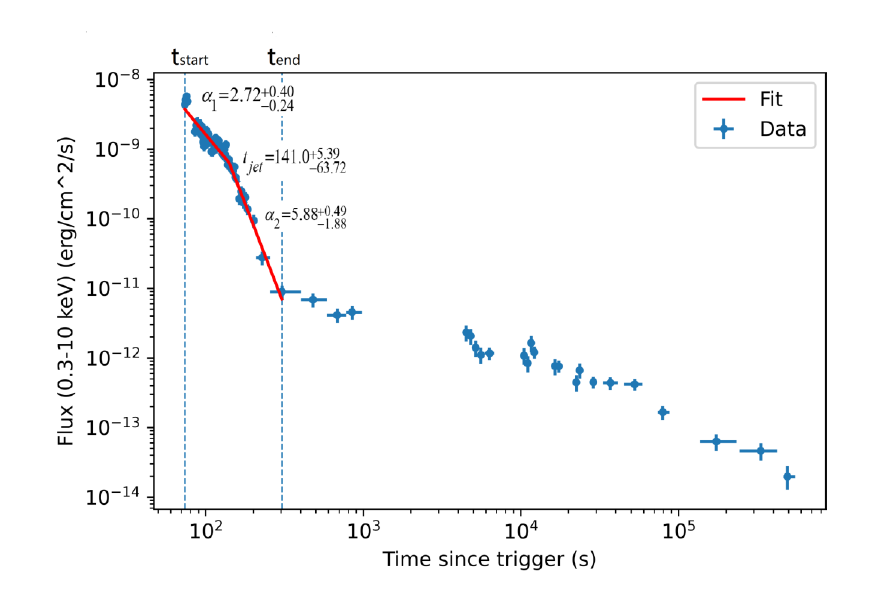}
    \caption{The X-ray afterglow fitting results for GRB 100425A. The scatter plot represents the afterglow flux light curve. The blue vertical dashed lines indicate the selected time window, and the red line corresponds to the broken power-law fitting result.}
    \label{fig_example}
\end{figure} 
We take GRB 100425A as an example, which is shown in Figure \ref{fig_example}. Data within the time of 70-400 s are selected for fitting. $t_{\rm start}$ and $t_{\rm end}$ respectively represent the times corresponding to the first and last data points within the selected time window. 
For GRB 100425A, $t_{\rm start}$ = 73.7 s, $t_{\rm end}$ = 304.4 s. Monte Carlo Markov Chain (MCMC) is employed for parameter estimation. The prior distributions of the parameters are as follows: The decay index $\alpha_1$ follows a uniform distribution between 0 and 5, $\alpha_2$ follows a uniform distribution between 3 and 10, the jet break time $t_{\rm jet}$ has a flat prior within the selected time window. 
For each GRB, the number of walkers is set to 100, and the MCMC chain length is 2000. 
The fitting results for GRB 100425A are $\alpha_1=2.72^{+0.40}_{-0.24}$, $\alpha_2=5.88^{+0.49}_{-1.88}$, $t_{\rm jet}=141.0^{+5.4}_{-63.7}$ s, which align with our expectations. 
Then, we can calculate $R_{\gamma} \theta_{\rm jet}^2$ using Equation (\ref{equ_1}) and constrain the jet opening angle. 
The calculation results for GRB 100425A show that its minimum jet opening angle $\theta_{\rm jet,min}=0.007$, its maximum jet opening angle $\theta_{\rm jet,max}=0.296$.
$\beta_1$ is the spectral index for the time interval same as for $\alpha_1$. The values of $\beta_1$ are taken from the {\it Swift}/XRT website, by imputing the time interval (73.7 s to 141.0 s) for time-sliced spectrum  \citep{Evans2009} \footnote{\url{https://www.swift.ac.uk/xrt_curves/allcurves.php}}. 
We get $\beta_1=2.58^{+0.14}_{-0.14}$.
If $\alpha_1 = \beta_1 +2$, we can say the emission at this time interval is the high latitude emission tail of the prompt $\gamma-$ray emission. 
However, for GRB 100425A, $\alpha_1 - \beta_1 \simeq 0.12$ make the $\alpha_1$ phase not likely from high  latitude emission. The jet angle constraint on this GRB is doubtful. 

The results for golden sample and silver sample are listed in Tables \ref{tab1} and \ref{tab2}, respectively. The golden sample includes the GRBs with a clear break and a followed steep decay. The silver sample includes the GRBs with either a smoother break or the steep decay phase, just contains a few observing data. 
Both of them should satisfy the closure relation, i.e., $\alpha_1=\beta_1+2$. Otherwise, the `normal' steep decay maybe does not come from the high latitude emission. Consequently, the follow-up decay maybe does not come from the jet break neither. 
To tolerate the possible errors, we select the GRBs with $\alpha_1-\beta_1 \in (1.5,2.5)$. 

For those GRBs outside this region, we also list them in Tables \ref{tab3} and \ref{tab4}, respectively. 
The violation of the closure relation might be that there is evolution of the spectrum, or the jet structure is not spherical. 
The break may still be caused by the jet edge. 
However, these constraints are not robustly reliable then.

{It is worth checking whether the break is from the changing of the spectral index $\beta_2$, which is measured during the period $(t_{\rm jet}, t_{\rm end})$. 
In this case, after the break, the light curve still obeys the closure relation, i.e., $\alpha_2 = \beta_2+2$. 
This is very unlikely for the golden samples as the breaking index $\alpha_2-\alpha_1$ is quite large, which means $\beta_2$ should be very large. 
For example, GRB 050814, with $\alpha_2-\alpha_1 \sim 1.21$, is the smallest in the golden sample. $\beta_2 = 1.58^{+0.19}_{-0.16}$ for earlier WT mode photons, and $\beta_2 = 1.00^{+0.20}_{-0.15}$ for later PC mode photons. 
However, $\beta_1 = 1.13^{+0.07}_{-0.07}$ is similar to $\beta_2$. It does not satisfy $\alpha_2 = \beta_2+2$. 
This could happen for the silver samples, as $\alpha_2-\alpha_1$ is not so large for some of them. 
GRBs 120219A and 171209A are the two examples, with $\alpha_2-\alpha_1$ being around 1.14 and 1.46 respectively. 
For 120219A, $\beta_2 = 2.01^{+0.14}_{-0.13}$ for earlier WT mode photons, and $\beta_2 = 1.02^{+0.13}_{-0.15}$ for later PC mode photons, while $\alpha_2 = 3.31^{+0.11}_{-0.11}$. They marginally follow the $\alpha_2 = \beta_2+2$. 
For GRB 171209A, there are only WT photons at this stage. $\beta_2 = 0.60^{+0.09}_{-0.09}$, while $\alpha_2= 3.32 ^{+0.16}_{-0.14}$. 
It is neither a confidentially high-latitude emission at a changed spectral segment. 
Therefore, though there are no decisive examples in the chosen data, one may be cautioned for those bursts with the even steeper decay not so steep. Checking the closure relation $\alpha_2 = \beta_2+2$ should be necessary before stating that the break is a jet break.
We list all the values of $\beta_2$ in all 4 tables. For GRBs with both WT and PC data, we only include $\beta_2$ of the WT data, as they are always closer to the break time and should be used to investigate the spectral crossing effect.
}

In addition to GRBs with not enough data, we do not include GRBs with the following features:
1. The dropdown decay index is not steep enough ($\alpha_2 < 3$)  (e.g., GRB 080721) because the decay itself might be caused by high-latitude emission. 
2. The $\alpha_1$ phase does not follow a single power law decay (e.g., GRB 060202) because it is very likely not from high latitude emission.
3. The steep decay is followed by a clear flare or flares clearly dominate the whole X-ray emissions (e.g., GRB 121031A) because the jet break might be covered by the flares, even if it exists. 
4. The dropdown decay exists, while the high latitude emission is missing (e.g., GRB 060428B). In this case, upper limit can be obtained as the jet break time should be smaller than the $t_{\rm start}$.

In table \ref{tab3}, for fitting the spectral index $\beta_1$ of GRB 210725A, two data modes (WT mode and PC mode) are contained in that time interval. 
$\beta_1=0.5^{+0.4}_{-0.3}$ and $0.28^{+0.25}_{-0.16}$, and the corresponding observed flux are around $1.9 \times 10^{-10} {\rm erg\, cm^{-2} \, s^{-1}}$ and $9.5 \times 10^{-11} {\rm erg\, cm^{-2} \, s^{-1}}$, for WT mode data and PC mode data, respectively. 
As the WT mode data slightly dominate the whole time interval, we choose the spectral index of the WT mode. 
For all other selected GRBs, there is no such issue, as the data are either only in WT mode or are clearly dominating by  WT mode. 

From Tables \ref{tab1} and \ref{tab2}, one can see the combination of radius and jet opening angle $R_{\gamma} \theta_{\rm jet}^2$ is quite concentrating at $10^{13}$ cm. 
The distribution is $(1.11\pm 0.35) \times 10^{13}$ cm for the golden sample, where the error is the standard deviation,  $(1.28\pm 0.17) \times 10^{13}$ cm for the silver sample, and  $(1.10 \pm 0.52) \times 10^{13}$ cm for both of them. 
Considering Eq. (\ref{equ_1}), $R_{\gamma} \theta_{\rm jet}^2$ is actually directly related to $t_{\rm jet}$. Therefore, the distribution of $t_{\rm jet}$ is tightly concentrated. 
However, this concentration is likely because of the selection effect. For shorter or longer $t_{\rm jet}$, the break is likely covered by the X-ray flares or afterglow emissions. 

Notice, in the whole work, we did not consider the redshift of each GRB. 
The reason is that only very few of them have redshift observations. 
Only 3 out of 12 GRBs in the whole golden and silver samples, while these 3 GRBs all have only photometric redshifts. Those are 
$z=5.77$ (photometric) for GRB 050814 \citep{2008A&A...490.1047C}, $z=2.03$ (photometric) for GRB 081230 \citep{2011A&A...526A.153K}, and $z=3.0$ (photometric) for GRB 090404 \citep{2013ApJ...778..172P}.
If the redshift is considered for these 4 GRBs, the concentration of $R_{\gamma} \theta_{\rm jet}^2$ is looser, which makes the tight distribution being intrinsic less likely. 

As there are no robust methods to constrain the radius $R_{\gamma}$, the range of $\theta_{\rm jet}$ is loosely constrained, consequently. It can be seen in the last two columns of the tables.

\begin{table}
    \centering
    \renewcommand{\arraystretch}{1.5}
    \begin{tabular}{c|c|c|c|c|c|c|c|c|c|c}
        \hline
        GRB  & $t_{\rm start}$ (s) & $t_{\rm end}$ (s) & $\alpha_1$ & $\beta_1$ & $t_{\rm jet}$ (s) & $\alpha_2$ & $\beta_2$ & $R_{\gamma} \theta_{\rm jet}^2$ (cm) & $\theta_{\rm jet, \min}$ &  $\theta_{\rm jet, \max}$ \\ \hline
         050814 & 166.6 & 841.4 & $2.73^{+0.17}_{-0.18}$ & $1.13^{+0.07}_{-0.07}$ & $325.6^{+32.7}_{-139.5}$ & $3.94^{+0.38}_{-0.72}$ & $1.58^{+0.19}_{-0.16}$ & $1.95^{+0.20}_{-0.84} \times 10^{13}$ & 0.011 & 0.464 \\
         081230 & 131.3 & 262.1 & $3.06^{+1.44}_{-1.92}$ & $0.44^{+0.26}_{-0.24}$ & $136.3^{+2.6}_{-2.0}$ & $7.07^{+0.24}_{-0.20}$ & $2.11^{+0.12}_{-0.12}$ & $8.18^{+0.16}_{-0.12} \times 10^{12}$ & 0.009 & 0.289 \\
         090404 & 92.0 & 196.7 & $2.86^{+0.29}_{-0.28}$ & $1.22^{+0.10}_{-0.09}$ & $121.5^{+1.4}_{-1.3}$ & $8.80^{+0.17}_{-0.18}$ & $2.31^{+0.10}_{-0.10}$ & $7.29^{+0.09}_{-0.08} \times 10^{12}$ & 0.008 & 0.272 \\
         150428B & 102.0 & 539.0 & $2.23^{+0.13}_{-0.12}$ & $0.91^{+0.07}_{-0.07}$ & $170.4^{+3.6}_{-3.1}$ & $4.87^{+0.12}_{-0.11}$ & $1.30^{+0.12}_{-0.08}$ & $10.02^{+0.02}_{-0.02} \times 10^{12}$ & 0.010 & 0.317 \\
         210818A & 53.3 & 146.9 & $1.33^{+0.10}_{-0.10}$ & $-0.35^{+0.07}_{-0.07}$ & $119.5^{+1.3}_{-4.7}$ & $8.66^{+0.77}_{-1.51}$ & $0.67^{+0.26}_{-0.23}$ & $7.17^{+0.08}_{-0.28} \times 10^{12}$ & 0.008 & 0.269 \\
         230420A & 81.0 & 300.6 & $2.55^{+1.24}_{-0.92}$ & $0.56^{+0.10}_{-0.09}$ & $100.3^{+63.3}_{-6.6}$ & $4.10^{+0.15}_{-0.29}$ & $0.61^{+0.10}_{-0.10}$ & $6.02^{+3.80}_{-0.40} \times 10^{12}$ & 0.007 & 0.313 \\
         231129A & 102.1 & 281.2 & $2.87^{+0.23}_{-0.25}$ & $0.47^{+0.15}_{-0.14}$ & $140.7^{+2.8}_{-3.0}$ & $6.09^{+0.13}_{-0.12}$ & $1.17^{+0.09}_{-0.09}$ & $8.44^{+0.17}_{-0.18} \times 10^{12}$ & 0.009 & 0.293 \\ \hline
    \end{tabular}
    \caption{Fitting results of the selected golden sample of {\it Swift}/XRT GRBs. $t_{\rm start}$ and $t_{\rm end}$ are the starting time and ending time choosing for the light curve fitting, respectively. $\alpha_1$ is the decay index before the break. $\beta_1$ is the spectral index of the X-ray emission during $t_{\rm start}$ and $t_{\rm jet}$, which is taken from the XRT website.  $t_{\rm jet}$ is the breaking time. $\alpha_2$ is the decay index after the break. {$\beta_2$ is the spectral index of the time interval between $t_{\rm jet}$ and $t_{\rm end}$.} $\theta_{\rm jet, \min}$ and  $\theta_{\rm jet, \max}$ are the minimum and maximum opening angle of the jet based on the lower limit and upper limit estimated from the $R_{\gamma}$. Same as the spectral-fit values from the official data center \citep{Evans2009}, the other errors are also in 90\% confidence.}
    \label{tab1}
\end{table}

\begin{table}
    \centering
    \renewcommand{\arraystretch}{1.5}
    \begin{tabular}{c|c|c|c|c|c|c|c|c|c|c}
        \hline
        GRB  & $t_{\rm start}$ (s) & $t_{\rm end}$ (s) & $\alpha_1$ & $\beta_1$ & $t_{\rm jet}$ (s) & $\alpha_2$ & $\beta_2$ & $R_{\gamma} \theta_{\rm jet}^2$ (cm) & $\theta_{\rm jet, \min}$ &  $\theta_{\rm jet, \max}$ \\ \hline
       080503 & 82.1 & 707.6 & $1.84^{+0.06}_{-0.05}$ & $0.16^{+0.05}_{-0.05}$ & $222.9^{+5.9}_{-6.1}$ & $5.31^{+0.18}_{-0.14}$ & $1.12^{+0.23}_{-0.21}$ & $1.34^{+0.04}_{-0.04} \times 10^{13}$ & 0.011 & 0.371 \\
       080613B & 77.4 & 258.2 & $2.09^{+0.09}_{-0.09}$ & $0.01^{+0.08}_{-0.07}$ & $181.6^{+2.2}_{-2.7}$ & $9.90^{+0.07}_{-0.17}$ & $0.60^{+0.28}_{-0.21}$ & $1.09^{+0.01}_{-0.02} \times 10^{13}$ & 0.010 & 0.332 \\
       120219A & 119.3 & 582.9 & $2.17^{+0.89}_{-0.78}$ & $0.20^{+0.50}_{-0.40}$ & $146.9^{+92.9}_{-9.8}$ & $3.31^{+0.11}_{-0.11}$ & $2.01^{+0.14}_{-0.13}$ & $8.81^{+5.57}_{-0.59} \times 10^{12}$ & 0.009 & 0.379 \\
       160313A & 158.9 & 571.7 & $1.92^{+0.42}_{-0.50}$ & $0.30^{+0.50}_{-0.40}$ & $276.9^{+31.5}_{-35.0}$ & $4.29^{+0.51}_{-0.45}$ & $1.80^{+0.70}_{-0.60}$ & $1.66^{+0.19}_{-0.21} \times 10^{13}$ & 0.012 & 0.430 \\
       171209A & 93.9 & 452.9 & $1.84^{+0.05}_{-0.05}$ & $0.27^{+0.05}_{-0.05}$ & $234.6^{+7.5}_{-8.2}$ & $3.32^{+0.16}_{-0.14}$ & $0.60^{+0.09}_{-0.09}$ & $1.41^{+0.04}_{-0.05} \times 10^{13}$ & 0.012 & 0.381 \\
    \hline
    \end{tabular}
    \caption{Fitting results of the selected silver sample of {\it Swift}/XRT GRBs. The parameters are the same as Table \ref{tab1}.}
    \label{tab2}
\end{table}

\begin{table}
    \centering
    \renewcommand{\arraystretch}{1.5}
    \begin{tabular}{c|c|c|c|c|c|c|c|c|c|c}
        \hline
        GRB  & $t_{\rm start}$ (s) & $t_{\rm end}$ (s) & $\alpha_1$ & $\beta_1$ & $t_{\rm jet}$ (s) & $\alpha_2$ & $\beta_2$ & $R_{\gamma} \theta_{\rm jet}^2$ (cm) & $\theta_{\rm jet, \min}$ &  $\theta_{\rm jet, \max}$ \\ \hline
         050724 & 80.3 & 471.5 & $1.68^{+0.38}_{-0.05}$ & $0.65^{+0.05}_{-0.05}$ & $201.7^{+59.3}_{-2.9}$ &$5.41^{+1.72}_{-0.13}$ & $1.56^{+0.14}_{-0.13}$ & $1.21^{+0.36}_{-0.02} \times 10^{13}$ & 0.011 & 0.396 \\
         060502A & 86.6 & 206.9 & $2.61^{+0.25}_{-0.30}$ & $2.40^{+0.27}_{-0.24}$ & $149.2^{+29.3}_{-33.4}$ & $3.87^{+1.61}_{-0.72}$ & $1.3^{+0.4}_{-0.3}$ & $8.95^{+1.76}_{-2.01} \times 10^{12}$ & 0.008 & 0.327 \\
         081221 & 73.3 & 298.7 & $1.26^{+0.29}_{-0.19}$ & $0.88^{+0.09}_{-0.08}$ & $107.6^{+2.7}_{-1.7}$ & $1.20^{+0.04}_{-0.04}$ & $1.47^{+0.05}_{-0.05}$ & $6.45^{+0.16}_{-0.10} \times 10^{12}$ & 0.008 & 0.257 \\
         100425A & 73.7 & 304.4 & $2.72^{+0.40}_{-0.24}$ & $2.58^{+0.14}_{-0.14}$ & $141.0^{+5.4}_{-63.7}$ &$5.88^{+0.49}_{-1.88}$ & $4.1^{+0.6}_{-0.6}$ & $8.46^{+0.32}_{-3.82} \times 10^{12}$ & 0.007 & 0.296 \\
         101030A & 72.9 & 224.2 & $2.11^{+0.24}_{-0.28}$ & $1.72^{+0.15}_{-0.15}$ & $109.7^{+5.3}_{-7.4}$ & $3.95^{+0.24}_{-0.21}$ & $1.84^{+0.21}_{-0.19}$ & $6.58^{+0.32}_{-0.44} \times 10^{12}$ & 0.008 & 0.248 \\
         101219B & 138.7 & 589.3 & $0.62^{+0.06}_{-0.06}$ & $0.46^{+0.04}_{-0.03}$ & $409.1^{+9.5}_{-11.3}$ & $4.10^{+0.43}_{-0.42}$ & $1.11^{+0.10}_{-0.09}$ & $2.45^{+0.06}_{-0.07} \times 10^{13}$ & 0.015 & 0.501 \\
         111123A & 94.3 & 787.5 & $1.20^{+0.01}_{-0.02}$ & $0.44^{+0.03}_{-0.03}$ & $542.0^{+3.7}_{-5.0}$ & $4.10^{+0.43}_{-0.42}$ & $1.40^{+0.11}_{-0.11}$ & $2.45^{+0.06}_{-0.07} \times 10^{13}$ & 0.015 & 0.501 \\
         120213A & 45.1 & 395.6 & $1.60^{+0.06}_{-0.06}$ & $1.69^{+0.07}_{-0.07}$ & $128.1^{+3.0}_{-3.4}$ & $4.07^{+0.08}_{-0.08}$ & $2.86^{+0.11}_{-0.10}$ & $7.68^{+0.18}_{-0.20} \times 10^{12}$ & 0.009 & 0.280 \\
         120308A & 120.6 & 299.7 & $2.58^{+0.10}_{-0.13}$ & $2.09^{+0.10}_{-0.09}$ & $220.6^{+5.8}_{-8.1}$ & $5.79^{+0.64}_{-0.71}$ & $2.9^{+0.5}_{-0.4}$ & $1.32^{+0.03}_{-0.05} \times 10^{13}$ & 0.011 & 0.367 \\
         151228B & 41.4 & 269.3 & $1.36^{+0.07}_{-0.07}$ & $1.42^{+0.09}_{-0.09}$ & $137.7^{+4.3}_{-3.8}$ & $4.28^{+0.27}_{-0.25}$ & $2.36^{+0.34}_{-0.26}$ & $8.26^{+0.26}_{-0.23} \times 10^{12}$ & 0.009 & 0.292 \\
         160501A & 120.1 & 1110.4 & $2.67^{+0.12}_{-0.17}$ & $1.90^{+0.16}_{-0.15}$ & $255.3^{+19.1}_{-80.5}$ & $3.84^{+0.26}_{-0.45}$ & $2.4^{+0.6}_{-0.5}$ & $1.53^{+0.11}_{-0.48} \times 10^{13}$ & 0.010 & 0.405 \\
         160804A & 138.4 & 895.2 & $1.45^{+0.02}_{-0.03}$ & $0.68^{+0.02}_{-0.02}$ & $519.3^{+3.2}_{-5.5}$ & $6.28^{+0.13}_{-0.18}$ & $1.39^{+0.07}_{-0.07}$ & $3.12^{+0.02}_{-0.03} \times 10^{13}$ & 0.018 & 0.560 \\
         170127A & 90.0 & 375.0 & $1.31^{+0.08}_{-0.09}$ & $0.48^{+0.11}_{-0.10}$ & $242.5^{+15.4}_{-13.1}$ & $3.66^{+0.62}_{-0.42}$ & $0.78^{+0.45}_{-0.27}$ & $1.45^{+0.09}_{-0.08} \times 10^{13}$ & 0.012 & 0.392 \\
         180314A & 150.9 & 382.9 & $2.28^{+0.17}_{-0.21}$ & $1.25^{+0.08}_{-0.04}$ & $264.9^{+55.1}_{-23.2}$ & $3.60^{+1.35}_{-0.45}$ & $1.33^{+0.14}_{-0.11}$ & $1.59^{+0.33}_{-0.14} \times 10^{13}$ & 0.012 & 0.438 \\
         180925A & 101.4 & 381.6 & $1.35^{+0.13}_{-0.12}$ & $0.38^{+0.10}_{-0.09}$ & $199.4^{+8.4}_{-6.9}$ & $3.74^{+0.22}_{-0.19}$ & $1.00^{+0.17}_{-0.16}$ & $1.20^{+0.05}_{-0.04} \times 10^{13}$ & 0.011 & 0.354 \\
         190219A & 91.6 & 320.7 & $2.73^{+0.19}_{-0.18}$ & $1.60^{+0.06}_{-0.06}$ & $139.2^{+2.7}_{-3.1}$ & $5.08^{+0.10}_{-0.10}$ & $1.87^{+0.09}_{-0.08}$ & $8.35^{+0.16}_{-0.19} \times 10^{12}$ & 0.009 & 0.292 \\
         200219A & 58.9 & 230.4 & $1.12^{+0.08}_{-0.08}$ & $0.27^{+0.07}_{-0.07}$ & $163.0^{+9.2}_{-6.7}$ & $3.12^{+0.25}_{-0.09}$ & $0.74^{+0.15}_{-0.14}$ & $9.78^{+0.55}_{-0.40} \times 10^{12}$ & 0.010 & 0.321 \\
         210421A & 82.7 & 404.5 & $2.06^{+1.93}_{-0.35}$ & $1.72^{+0.08}_{-0.08}$ & $111.4^{+62.4}_{-2.3}$ & $5.05^{+0.41}_{-0.10}$ & $1.87^{+0.11}_{-0.10}$ & $6.69^{+3.74}_{-1.36} \times 10^{12}$ & 0.007 & 0.323 \\
         210725A & 118.0 & 643.7 & $1.04^{+0.14}_{-0.14}$ & $0.5^{+0.4}_{-0.3}$ & $403.9^{+24.1}_{-24.8}$ & $5.04^{+1.01}_{-0.83}$ & $0.8^{+0.6}_{-0.5}$ & $2.42^{+0.14}_{-0.15} \times 10^{13}$ & 0.015 & 0.506 \\
    \hline
    \end{tabular}
    \caption{Similar to Table \ref{tab1}, but with sample that the closure relation $\alpha_1=\beta_1+2$ is not well satisfied.}
    \label{tab3}
\end{table}

\begin{table}
    \centering
    \renewcommand{\arraystretch}{1.5}
    \begin{tabular}{c|c|c|c|c|c|c|c|c|c|c}
        \hline
        GRB  & $t_{\rm start}$ (s) & $t_{\rm end}$ (s) & $\alpha_1$ & $\beta_1$ & $t_{\rm jet}$ (s) & $\alpha_2$ & $\beta_2$ & $R_{\gamma} \theta_{\rm jet}^2$ (cm) & $\theta_{\rm jet, \min}$ &  $\theta_{\rm jet, \max}$ \\ \hline
       071227 & 108.8 & 464.1 & $1.58^{+0.26}_{-0.25}$ & $0.51^{+0.16}_{-0.15}$ & $183.3^{+5.9}_{-5.8}$ & $4.90^{+0.39}_{-0.31}$ & $1.1^{+0.4}_{-0.4}$ & $1.10^{+0.04}_{-0.03} \times 10^{13}$ & 0.010 & 0.338 \\
       080123 & 110.4 & 549.0 & $1.82^{+0.13}_{-0.13}$ & $0.58^{+0.10}_{-0.09}$ & $214.0^{+4.8}_{-5.1}$ & $7.92^{+1.02}_{-0.93}$ & $0.8^{+0.5}_{-0.4}$ & $1.28^{+0.03}_{-0.03} \times 10^{13}$ & 0.011 & 0.362 \\
       100725A & 84.1 & 345.3 & $0.81^{+0.12}_{-0.12}$ & $0.01^{+0.09}_{-0.09}$ & $202.3^{+12.9}_{-9.6}$ & $3.13^{+0.24}_{-0.10}$ & $0.37^{+0.21}_{-0.20}$ & $1.21^{+0.08}_{-0.06} \times 10^{13}$ & 0.011 & 0.359 \\
       120326A & 52.9 & 232.0 & $2.88^{+0.13}_{-0.31}$ & $2.56^{+0.20}_{-0.19}$ & $103.6^{+7.9}_{-25.8}$ & $4.35^{+0.32}_{-0.45}$ & $2.8^{+0.4}_{-0.4}$ & $6.22^{+0.47}_{-1.55} \times 10^{12}$ & 0.007 & 0.259 \\
       160412A & 69.1 & 255.1 & $3.16^{+0.17}_{-0.17}$ & $1.87^{+0.14}_{-0.14}$ & $110.4^{+3.8}_{-2.9}$ & $5.87^{+0.19}_{-0.16}$ & $2.55^{+0.19}_{-0.18}$ & $6.62^{+0.23}_{-0.18} \times 10^{12}$ & 0.008 & 0.262 \\ 
       171007A & 76.7 & 374.4 & $0.59^{+0.21}_{-0.23}$ & $0.38^{+0.20}_{-0.18}$ & $170.7^{+7.5}_{-7.2}$ & $6.18^{+0.45}_{-0.38}$ & $2.7^{+2.8}_{-1.8}$ & $1.02^{+0.05}_{-0.04} \times 10^{13}$ & 0.010 & 0.327 \\
       181203A & 66.6 & 147.0 & $2.51^{+0.79}_{-0.93}$ & $2.33^{+1.05}_{-0.69}$ & $85.8^{+8.2}_{-3.3}$ & $4.52^{+0.66}_{-0.57}$ & $1.94^{+1.46}_{-0.72}$ & $5.15^{+0.49}_{-0.20} \times 10^{12}$ & 0.007 & 0.237 \\ 
       210912A & 49.7 & 109.4 & $2.02^{+0.85}_{-1.02}$ & $2.60^{+0.70}_{-0.50}$ & $65.8^{+9.1}_{-3.5}$ & $4.52^{+0.68}_{-0.52}$ & $2.59^{+1.15}_{-0.74}$ & $3.95^{+0.54}_{-0.21} \times 10^{12}$ & 0.006 & 0.212 \\
    \hline
    \end{tabular}
    \caption{Similar to Table \ref{tab2}, but with sample that the closure relation $\alpha_1=\beta_1+2$ is not well satisfied.}
    \label{tab4}
\end{table}

\section{Conclusion and discussion}\label{sec:Con}
In this work, we investigated the jet opening angles of Gamma-Ray Bursts (GRBs) by analyzing the steep decay phase and the dropdown decay phase observed in the early X-ray afterglow, which is attributed to high latitude emission and the edge of the jet, respectively. Using data from the {\it Swift}/XRT, we identified GRBs exhibiting a noticeable break in their light curves and employed a broken power-law model to characterize the decay before and after the break. 

Our analysis is based on the assumption that the emitting surface is spherical and isotropic in the comoving frame, and that the jet is oriented towards the observer. We constrained the radiation radius using both lower and upper limits based on the compactness problem and deceleration radius, respectively. The jet opening angle was then inferred from the breaking time, which marks the transition from steep to dropdown decay.

We applied this method to GRBs {with steep decays}, and used a Monte Carlo Markov Chain (MCMC) for parameter estimation. The results demonstrated that the jet opening angles could be constrained within a certain range, although the exact values are dependent on the assumed radiation radius. 
Once there are reliable ways to constrain the radiation radius, the jet opening angle can be well determined. {One should be cautioned that the even steeper decay phase ($\alpha_2$) being caused by spectral crossing. In this case, the closure relation $\alpha_2 = \beta_2 +2 $ still obeys, and the break cannot be used to constrain the jet opening angle.}


It is worth constraining $\Gamma$ and $\theta_{\rm jet}$ jointly with multi-method. 
Once the $\Gamma$ is determined, one may use it to determine the radius $R_\gamma$. 
Notice that the timescale of each GRB is easy to obtain, though the uncertainty is high and the diversity of each pulse high.
However, since there are different methods to determine $\Gamma$ \citep[see][and references therein]{2011ApJ...726L...2Z}, and different methods might be used for different GRBs, it is difficult to find a unified method to constrain all GRBs. 
Therefore, one may look into the details of each burst and try to find a proper method to constrain both the $\Gamma$ and $\theta_{\rm jet}$.

\section*{Acknowledgements}
We thank the anonymous referee for the insightful suggestions and comments. Y.-C. Z. thanks Rodolfo Barniol Duran for the valuable discussion. This work was supported by the National SKA Program of China (2022SKA0130100).
The computation was completed on the HPC Platform of Huazhong University of Science and Technology.




\bibliographystyle{aasjournal}
\bibliography{bib} 

\begin{thebibliography}{}
\expandafter\ifx\csname natexlab\endcsname\relax\def\natexlab#1{#1}\fi
\providecommand{\url}[1]{\href{#1}{#1}}
\providecommand{\dodoi}[1]{doi:~\href{http://doi.org/#1}{\nolinkurl{#1}}}
\providecommand{\doeprint}[1]{\href{http://ascl.net/#1}{\nolinkurl{http://ascl.net/#1}}}
\providecommand{\doarXiv}[1]{\href{https://arxiv.org/abs/#1}{\nolinkurl{https://arxiv.org/abs/#1}}}

\bibitem[{{Barniol Duran} \& {Kumar}(2009)}]{2009MNRAS.395..955B}
{Barniol Duran}, R., \& {Kumar}, P. 2009, \mnras, 395, 955, \dodoi{10.1111/j.1365-2966.2009.14584.x}

\bibitem[{{Bloom} {et~al.}(2003){Bloom}, {Frail}, \& {Kulkarni}}]{Bloom2003}
{Bloom}, J.~S., {Frail}, D.~A., \& {Kulkarni}, S.~R. 2003, \apj, 594, 674, \dodoi{10.1086/376961}

\bibitem[{{Curran} {et~al.}(2008){Curran}, {Wijers}, {Heemskerk}, {Starling}, {Wiersema}, \& {van der Horst}}]{2008A&A...490.1047C}
{Curran}, P.~A., {Wijers}, R.~A.~M.~J., {Heemskerk}, M.~H.~M., {et~al.} 2008, \aap, 490, 1047, \dodoi{10.1051/0004-6361:200810545}

\bibitem[{{Daigne} \& {Mochkovitch}(1998)}]{Daigne1998}
{Daigne}, F., \& {Mochkovitch}, R. 1998, \mnras, 296, 275, \dodoi{10.1046/j.1365-8711.1998.01305.x}

\bibitem[{{Evans} {et~al.}(2009){Evans}, {Beardmore}, {Page}, {Osborne}, {O'Brien}, {Willingale}, {Starling}, {Burrows}, {Godet}, {Vetere}, {Racusin}, {Goad}, {Wiersema}, {Angelini}, {Capalbi}, {Chincarini}, {Gehrels}, {Kennea}, {Margutti}, {Morris}, {Mountford}, {Pagani}, {Perri}, {Romano}, \& {Tanvir}}]{Evans2009}
{Evans}, P.~A., {Beardmore}, A.~P., {Page}, K.~L., {et~al.} 2009, \mnras, 397, 1177, \dodoi{10.1111/j.1365-2966.2009.14913.x}

\bibitem[{{Fenimore} {et~al.}(1993){Fenimore}, {Epstein}, \& {Ho}}]{1993A&AS...97...59F}
{Fenimore}, E.~E., {Epstein}, R.~I., \& {Ho}, C. 1993, \aaps, 97, 59

\bibitem[{{Frail} {et~al.}(2001){Frail}, {Kulkarni}, {Sari}, {Djorgovski}, {Bloom}, {Galama}, {Reichart}, {Berger}, {Harrison}, {Price}, {Yost}, {Diercks}, {Goodrich}, \& {Chaffee}}]{Frail2001}
{Frail}, D.~A., {Kulkarni}, S.~R., {Sari}, R., {et~al.} 2001, \apjl, 562, L55, \dodoi{10.1086/338119}

\bibitem[{{Gehrels} {et~al.}(2004){Gehrels}, {Chincarini}, {Giommi}, {Mason}, {Nousek}, {Wells}, {White}, {Barthelmy}, {Burrows}, {Cominsky}, {Hurley}, {Marshall}, {M{\'e}sz{\'a}ros}, {Roming}, {Angelini}, {Barbier}, {Belloni}, {Campana}, {Caraveo}, {Chester}, {Citterio}, {Cline}, {Cropper}, {Cummings}, {Dean}, {Feigelson}, {Fenimore}, {Frail}, {Fruchter}, {Garmire}, {Gendreau}, {Ghisellini}, {Greiner}, {Hill}, {Hunsberger}, {Krimm}, {Kulkarni}, {Kumar}, {Lebrun}, {Lloyd-Ronning}, {Markwardt}, {Mattson}, {Mushotzky}, {Norris}, {Osborne}, {Paczynski}, {Palmer}, {Park}, {Parsons}, {Paul}, {Rees}, {Reynolds}, {Rhoads}, {Sasseen}, {Schaefer}, {Short}, {Smale}, {Smith}, {Stella}, {Tagliaferri}, {Takahashi}, {Tashiro}, {Townsley}, {Tueller}, {Turner}, {Vietri}, {Voges}, {Ward}, {Willingale}, {Zerbi}, \& {Zhang}}]{Gehrels2004}
{Gehrels}, N., {Chincarini}, G., {Giommi}, P., {et~al.} 2004, \apj, 611, 1005, \dodoi{10.1086/422091}

\bibitem[{{Hasco{\"e}t} {et~al.}(2012){Hasco{\"e}t}, {Daigne}, \& {Mochkovitch}}]{2012A&A...542L..29H}
{Hasco{\"e}t}, R., {Daigne}, F., \& {Mochkovitch}, R. 2012, \aap, 542, L29, \dodoi{10.1051/0004-6361/201219339}

\bibitem[{{Kr{\"u}hler} {et~al.}(2011){Kr{\"u}hler}, {Schady}, {Greiner}, {Afonso}, {Bottacini}, {Clemens}, {Filgas}, {Klose}, {Koch}, {K{\"u}pc{\"u}-Yolda{\c{s}}}, {Oates}, {Olivares E.}, {Page}, {McBreen}, {Nardini}, {Nicuesa Guelbenzu}, {Rau}, {Roming}, {Rossi}, {Updike}, \& {Yolda{\c{s}}}}]{2011A&A...526A.153K}
{Kr{\"u}hler}, T., {Schady}, P., {Greiner}, J., {et~al.} 2011, \aap, 526, A153, \dodoi{10.1051/0004-6361/201015327}

\bibitem[{{Kumar} \& {Panaitescu}(2000)}]{2000ApJ...541L..51K}
{Kumar}, P., \& {Panaitescu}, A. 2000, \apjl, 541, L51, \dodoi{10.1086/312905}

\bibitem[{{Kumar} \& {Zhang}(2015)}]{2015PhR...561....1K}
{Kumar}, P., \& {Zhang}, B. 2015, \physrep, 561, 1, \dodoi{10.1016/j.physrep.2014.09.008}

\bibitem[{{Lithwick} \& {Sari}(2001)}]{Lithwick2001}
{Lithwick}, Y., \& {Sari}, R. 2001, \apj, 555, 540, \dodoi{10.1086/321455}

\bibitem[{{Perley} \& {Perley}(2013)}]{2013ApJ...778..172P}
{Perley}, D.~A., \& {Perley}, R.~A. 2013, \apj, 778, 172, \dodoi{10.1088/0004-637X/778/2/172}

\bibitem[{{Piran}(1999)}]{1999PhR...314..575P}
{Piran}, T. 1999, \physrep, 314, 575, \dodoi{10.1016/S0370-1573(98)00127-6}

\bibitem[{{Piran}(2004)}]{2004RvMP...76.1143P}
---. 2004, Reviews of Modern Physics, 76, 1143, \dodoi{10.1103/RevModPhys.76.1143}

\bibitem[{{Rees} \& {Meszaros}(1994)}]{Rees1994}
{Rees}, M.~J., \& {Meszaros}, P. 1994, \apjl, 430, L93, \dodoi{10.1086/187446}

\bibitem[{{Rhoads}(1999)}]{Rhoads1999}
{Rhoads}, J.~E. 1999, \apj, 525, 737, \dodoi{10.1086/307907}

\bibitem[{{Sari} {et~al.}(1999){Sari}, {Piran}, \& {Halpern}}]{Sari1999}
{Sari}, R., {Piran}, T., \& {Halpern}, J.~P. 1999, \apjl, 519, L17, \dodoi{10.1086/312109}

\bibitem[{{Zhang}(2018)}]{2019pgrb.book.....Z}
{Zhang}, B. 2018, {The physics of gamma-ray bursts} (Cambridge University Press)

\bibitem[{{Zhang} {et~al.}(2006){Zhang}, {Fan}, {Dyks}, {Kobayashi}, {M{\'e}sz{\'a}ros}, {Burrows}, {Nousek}, \& {Gehrels}}]{Zhang2006}
{Zhang}, B., {Fan}, Y.~Z., {Dyks}, J., {et~al.} 2006, \apj, 642, 354, \dodoi{10.1086/500723}

\bibitem[{{Zou} {et~al.}(2011){Zou}, {Fan}, \& {Piran}}]{2011ApJ...726L...2Z}
{Zou}, Y.-C., {Fan}, Y.-Z., \& {Piran}, T. 2011, \apjl, 726, L2, \dodoi{10.1088/2041-8205/726/1/L2}

\end{thebibliography}

\end{document}